# Asymmetrical precipitation on {10-12} twin boundary in magnesium alloy

Xin-Fu Gu[*], Man Wang, Zhang-Zhi Shi, Leng Chen, Ping Yang

School of Materials Science and Engineering, University of Science and Technology Beijing, Beijing, 100083, China

*Corresponding author: Xin-Fu Gu, xinfugu@ustb.edu.cn

**Abstract**

Precipitation on deformation defects is essential for enhancing mechanical properties of age-hardenable alloys. {10-12} twinning is common in deformed alloys with hexagonal-close-packed structure. In this work, we revealed how {10-12} twin boundary (TB) influences $\beta$-$Mg_{17}Al_{12}$ precipitation in Mg-9Al-1Zn alloy. The precipitates on the TB are rod-like, while others are lath-shape. The precipitates hold the Burgers orientation relationship (OR) only with twin or matrix, contrary to traditional wisdom in others alloys that precipitates on TB symmetrically keep an OR with both twin and matrix. Moreover, certain precipitate variants are absent, and a new rule for variant selection on TB is proposed.

Combination of pre-deformation and ageing process has shown be an effective way to strengthen magnesium alloys [1-5]. The deformation defects will act as heterogeneous nucleation sites and promote the precipitation [6], and in turn these precipitates will pin the movement of defects, and improve the mechanical property [7, 8]. Similar to other alloy systems, the precipitates in magnesium alloy also often exhibit preferred crystallographic features [9], such as specific orientation relationship (OR), interfacial orientation, morphology, growth direction etc. These crystallographic features are one of the determinant factors of mechanical property [10-13]. Therefore, the understanding of precipitation crystallography at deformation defects in magnesium is indispensable to control the mechanical property, however, the related studies are limited.

The crystallographic restriction of deformation defects on precipitation has been studied in other alloy systems mostly with cubic matrix due to the importance of steel, titanium alloy etc. Generally speaking, some particular variants, which holds crystallographically equivalent OR with respect to the matrix, would be preferred on the deformation defects, i.e. variant selection happens [14-18]. The geometry criterion for the variant selection on dislocations is that the largest misfit direction between the variants and matrix is close to the Burgers vector of the dislocation so that the transformation strain can be effectively accommodated [14-17], while the criterion for the variant selection on twin boundary (TB) is that the OR between the precipitate and either side of TB is close to that precipitated in matrix resulting in low interphase boundary energies at both side of TB, such as Cr precipitates (bcc, body-centered cubic structure) in Ni matrix (fcc, face-centered cubic structure) [15], γ austenite (fcc) precipitates in δ ferrite matrix (bcc) [19] and α phase (hcp, hexagonal-close-packed structure) precipitates in β matrix (bcc) [18]. However, the OR in these cases should be exactly or nearly symmetry around the twin boundary, but generally an OR will not satisfy this requirement. In this work, we will show such a general case and might shed some light on the effect of defects in hcp structure on precipitation crystallography.

In magnesium alloys, twinning is an important deformation mode to accommodate deformation strain due to limited slip systems. Since {10-12}<10-11> extension twinning mode is not only common in Mg alloys, but also common in other HCP metals, precipitation on {10-12} twin boundary is mainly focused here. In this work, the precipitation crystallography of β-$Mg_{17}Al_{12}$ (space group I-43m, a = 1.056nm [20]) on {10-12} TB in widely used Mg-Al-Zn alloy will be characterized by transmission electron microscopy (TEM) and also be statically investigated by transmission electron backscatter diffraction (t-EBSD) or transmission Kikuchi diffraction (TKD). Finally, a novel variant selection rule different from other alloy systems will be reported.

As-casted Mg-9Al-1Zn (AZ91) alloy was homogenized at 415°C for 24 h and quenched by water. Compression samples with a size of ϕ10×15mm were cut from the solution heat-treated sample, then compressed at room temperature with ~4% strain to introduce proper amount of deformation twins. Finally, the samples were aged at 300°C for 2-6 h. The method for preparing TEM specimens was the same as in our previous work [21]. The TEM observation was carried out with Talos F200 (200kV, FEI) and the statistic study of the crystal orientation was based on EBSD (Oxford Instrument). Both orientation analysis and pole figures were made by our recently developed software PTCLab [22].

Figure 1(a-b) shows the morphology of precipitates along TB and in matrix after ageing at 300°C for 2 h and 6 h respectively. The white contrast in the image arises from precipitates, while the dark background is the matrix. The precipitates in matrix have a lath-like shape, while the precipitates on TB are nodules. After deep etching, three-dimensional morphology could be clearly seen in Figure 1(c). The lath-shaped precipitates in matrix has a large length/thickness ratio, and their different inclinations are due to different variants, consistent with a previous study [20]. The precipitates on TB are actually with rod shapes. Based on the composition by energy dispersive spectroscopy in TEM shown in Figure 1(d), both the precipitates in the matrix and those on TB are β-$Mg_{17}Al_{12}$ type precipitates. Therefore, the deformation TB modifies the morphology of β-$Mg_{17}Al_{12}$ precipitates from plate shape to rod shape.

Figure 2 shows precipitation crystallography of β-$Mg_{17}Al_{12}$ in matrix after ageing at 300°C for 2h. The precipitates in Figure 2(a) are viewed with their habit planes edge-on along $[2\text{-}1\text{-}10]_\alpha$ zone axis in α-Mg matrix. A high resolution image near the habit plane (HP) is shown in Figure 2(b). The periodic strain contrasts in the interface are caused by interfacial dislocations with its spacing about 7.5 nm. Figure 2(c) is the Fast Fourier Transformation (FFT) of Figure 2(b), based on which we could find that the OR between β-$Mg_{17}Al_{12}$ and the α-Mg matrix follows the Burgers OR $(0001)_\alpha$ // $(011)_\beta$ and $[2\text{-}1\text{-}10]_\alpha$ // $[1\text{-}11]_\beta$ with the HP paralleling to $(0001)_\alpha$ // $(011)_\beta$. The Burgers OR and the HP are consistent with previous reports [23-27].

Figure 3 shows the precipitation on TB. In Figure 3(a-b), the twin boundary $(0\text{-}112)_\alpha$ is edge-on viewed along $[2\text{-}1\text{-}10]_\alpha$ zone axis. The precipitate could grow across the TB as in Figure 3(a) or at one side of TB as in Figure 3(b). It is noted that the main facets of the precipitates are not edge-on at this $[2\text{-}1\text{-}10]_\alpha$ zone axis, though the TB is edge-on. From other $<2\text{-}1\text{-}10>_\alpha$ zone axis, main facets could be edge-on as shown in Figure 3(c-f), such as the main facets F1 and F2 of the precipitate in Figure 3(d) is edge on. Figure 3(f) is the FFT of Figure 3(e). Accordingly, the zone axis is indexed as $[2\text{-}1\text{-}10]_\alpha$ // $[1\text{-}11]_\beta$ and the close packed planes are also parallel to each other, i.e. $(0001)_\alpha$ // $(011)_\beta$, so the OR of the precipitates at TB also follows Burgers OR. In TB, the matrix and twin shares one common $<2\text{-}1\text{-}10>_\alpha$ (Figure 4b), and this common $<2\text{-}1\text{-}10>_\alpha$ is expected to be utilized to define the Burgers OR if the precipitates could keep the Burgers OR with both twin and matrix. However, it is interesting to note that the $<2\text{-}1\text{-}10>_\alpha$ // $<1\text{-}11>_\beta$ direction used to define the Burgers OR is not lying in the TB, i.e. the precipitates is probably asymmetrically precipitated at TB.

In order to confirm this asymmetrical phenomenon, the OR of the precipitates at TB is statistically studied by transmission EBSD or TKD, and the result is shown in Figure 4. Figure 4(a) shows the band contrast measured by EBSD, and precipitates along the TB are numbered for further analysis. The twin in Figure 4(a) is $(0\text{-}112)_\alpha$ twin according to Figure 4(b). The orientations of nine precipitates in Figure 4(a) are shown in $<1\text{-}11>_\beta$ pole figure as Figure 4(c) and $\{011\}_\beta$ pole figure as Figure 4(d), respectively. In addition, the poles of $<2\text{-}1\text{-}10>_\alpha$ and $\{0001\}_\alpha$ poles of both twin and matrix are added respectively for convenience of determining OR. It could be seen from Figure 4(c-d), the precipitates only keep the Burgers OR with either matrix or twin. The common [2-1-10]$_\alpha$ in TB as indicated by two headed arrow in Figure 4(c) deviates from any <1-11>$_\beta$ poles. This result is reproducible based on the study over 50 precipitates, and another example is shown in Figure S3 in Supplementary materials for reference. In a word, the precipitates with <2-1-10>$_\alpha$ // <1-11>$_\beta$ lying in TB are systematically absent.

There are six variants of exact Burgers OR as listed in Table 1. Note that if the small deviation (0.3°) of Burgers OR [27] is taken into account, there would be 12 variants. However, such a small deviation makes no difference for our final conclusion, so the exact Burgers OR is considered here for simplicity. The common variant selection rule for precipitation on TB is that preferred precipitates keep the OR at both side of the TB as much as possible [15, 18, 19], named Rule I. The misorientation between the precipitates and twin is shown in last column of Table 1. Accordingly, among the six variants, if Burgers OR at both side of TB are approximately satisfied, only two variants, i.e. V2 and V4, are satisfied with a deviation of 8.2°. However, the precipitates with Nos. 1, 2, 4, 7-8 in Figure 4(a) violate this Rule I, since the {011}$_\beta$ poles deviate more than 20° from {0001}$_\alpha$ plane in twin as shown in Figure 4(d). According to our statistic result, only about 25% precipitates obey Rule I when considering a deviation angle of 10° from Burgers OR. Regarding the relationship between the twin shear {01-1-1}$_\alpha$ and the largest misfit direction of the transformation strain during precipitation, the largest misfit direction at Burgers OR is along <0001>$_\alpha$//<011>$_\beta$ based on the lattice correspondence in Ref. [28], and the angle between the twin shear and the largest misfit direction is fixed and same for all variants, therefore, the possible criterion that transformation strain is effectively accommodated by twin shear is not applicable here. However, we could draw lessons from the variant selection on grain boundary that the growth direction of preferred variant should get close to grain boundary to reduce the boundary area and thus reduce the total energy [29, 30]. In our case, the growth direction is close to <2-1-10>$_\alpha$//<1-11>$_\beta$ which defines Burgers OR [26-28], and this direction should lie in the TB base on the above rule, but it disagrees with the experimental result. Instead, the growth direction in our case largely deviate from the TB. Contrary to general grain boundary, the $\{10\text{-}12\}_\alpha$ TB is a low energy boundary with boundary energy about 120 mJ/m$^2$ [31], while the energy of general semi-coherent boundary is several times higher [32]. If the growth direction is close to TB, it will eliminate the TB area effectively, and cause the energy increment. This might be one of the reason for the variant selection in our case. Therefore, we propose a new rule for variant selection in TB if the variant could not keep the same OR at both side of TB, and the rule is that the growth direction of preferred variant should deviate from the TB as far as possible to minimize the possible elimination of the low energy boundary, named Rule II.

It is noted that ordered structure could form in segregated defects, such as in general boundary [33, 34], twin boundary [35, 36], and the relationship between the ordered structure and variant selection is worth to be further investigated. Nevertheless, the Burgers OR is a general OR in hexagonal systems, such as Mg-Al [23-26], Mg-Sn [37, 38], reverse transformation in titanium etc. The present result would be helpful for studying the effect of defect on phase transformation in magnesium alloy, or reverse transformation in titanium alloy etc.

In summary, the precipitation crystallography on TB is studied in an Mg-Al-Zn alloy. The morphology of precipitates will be changed from lath shape to rod shape by the crystallographic restriction of twin, while the orientation relationship between the precipitates and matrix/twin in both cases is Burgers OR. However, certain precipitate variants on TB are absent. For the preferred variants on TB, the <2-1-10>$_\alpha$ // <1-11>$_\beta$ direction defining in the Burgers OR does not lie in the TB. The precipitates only keeps the Burgers OR with one side of twin, contrary to conventional selection rules that the variant symmetrically shares the same OR at both side of TB as shown in other alloys. A new variant selection criterion for precipitation on TB is proposed, i.e. the growth direction of preferred variant should deviate the TB as far as possible to avoid the elimination of low energy TB during their growth.

**Acknowledgement**

This work was supported by Fundamental Research Funds for the Central Universities

**Appendix A. Supplementary data**

Tables

Table 1. Six variants of β-Mg$_{17}$Al$_{12}$ precipitates at Burgers orientation relationship (OR) in Mg matrix. The misorientation between different variants is shown in third column, and θ is the deviation angle from the Burgers OR with the other side of twin boundary (0-112)$_\alpha$.

Figures

Figure 1. SEM image of β-$Mg_{17}Al_{12}$ precipitates in Mg-Al-Zn alloy aged at 300°C for a) 2 h, b) 6h. c) Tilt view of the deep-etched sample aged for 2h. d) Element mapping of the precipitates by TEM-EDS. The twin boundaries (T. B.) are indicated by yellow lines. The precipitates on the twin boundaries and in matrix have different morphologies.

Figure 2. TEM images of β-$Mg_{17}Al_{12}$ precipitates in matrix in Mg-Al-Zn alloy aged at 300°C for 2h. a) Low-mag view of the precipitates along $[2-1-10]_\alpha$ zone axis of hcp (α) matrix, b) High resolution image of a single precipitate around its semi-coherent interface, where the periodic strain contrasts on the interface are caused by interfacial dislocations, c) Fast Fourier Transformation (FFT) of image (b), which shows the Burgers orientation relationship: $(0001)_\alpha$ // $(011)_\beta$ and $[2-1-10]_\alpha$ // $[1-11]_\beta$ between the precipitates and the matrix.

Figure 3. TEM images of β-$Mg_{17}Al_{12}$ precipitates along the twin boundary in Mg-Al-Zn alloy aged at 300°C for 2h. a) A precipitate across the twin boundary viewed along $[2-1-10]_\alpha$ zone axis with twin boundary (T. B.) edge-on, and the inset are the diffraction pattern across the twin, which shows $\{0-112\}_\alpha$ type twin, b) A precipitate nearly in one side of twin boundary viewed along $[2-1-10]_\alpha$ zone axis, c) HAADF-STEM image of two precipitates with their interface edge-on along $[-12-10]_\alpha$, d) enlarged view of one of the precipitate in figure (c), and the Facet 1 (F1) and Facet 2 (F2) are edge-on, c) High resolution image around F1, and the periodic strain contrasts on the interface are caused by interfacial dislocations, f) Fast Fourier Transformation (FFT) of image (e), which shows the Burgers orientation relationship.

Figure 4. Statistic study of the OR between the matrix and the precipitates along the twin boundary by transmission EBSD (t-EBSD). a) Band contrast (BC) mapping, and the numbers in the figure indicates the precipitates for further analysis, b) pole figure for $\{0-112\}_\alpha$ type twin in figure (a), c) $<1-11>_\beta$ pole figure for the precipitates

numbered in figure (a), and the <2-1-10>$_\alpha$ directions in matrix are also superimposed, d) {011}$_\beta$ pole figure of the precipitates numbered in Figure (a) and the {0001}$_\alpha$ planes are also superimposed. Blue circles indicate the indices in matrix while orange squares are for those in twin. For <1-11>$_\beta$ pole figure, there are four <1-11>$_\beta$ directions for one precipitate and they are connected with big circles for easier read, while in {011}$_\beta$ pole figure, there are six poles of {011}$_\beta$ planes for one precipitate, and they are also connected with big circles.

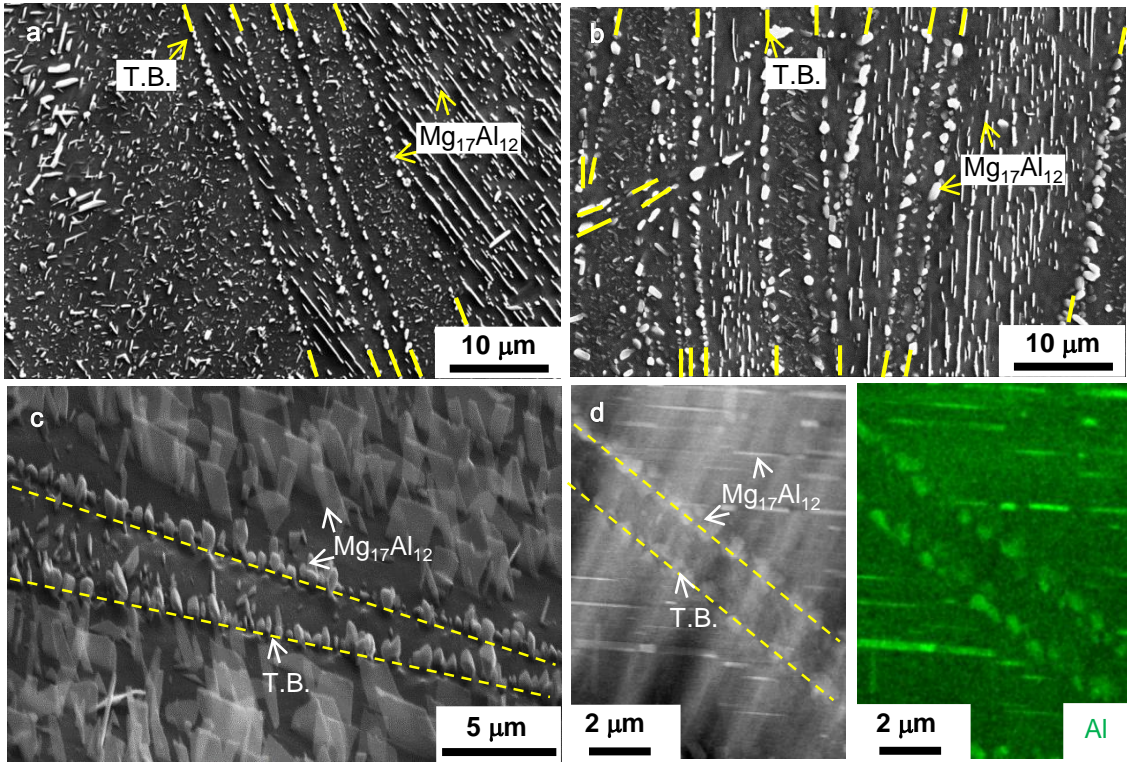

Figure 1. SEM image of β-Mg$_{17}$Al$_{12}$ precipitates in Mg-Al-Zn alloy aged at 300°C for a) 2 h, b) 6h. c) Tilt view of the deep-etched sample aged for 2h. d) Element mapping of the precipitates by TEM-EDS. The twin boundaries (T. B.) are indicated by yellow lines. The precipitates on the twin boundaries and in matrix have different morphologies.

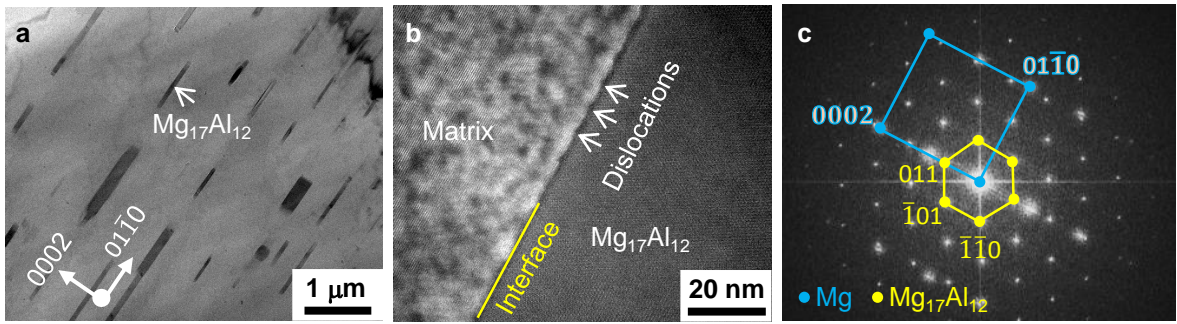

Figure 2. TEM images of β-$Mg_{17}Al_{12}$ precipitates in matrix in Mg-Al-Zn alloy aged at 300°C for 2h. a) Low-mag view of the precipitates along $[2\text{-}1\text{-}10]_\alpha$ zone axis of hcp (α) matrix, b) High resolution image of a single precipitate around its semi-coherent interface, where the periodic strain contrasts on the interface are caused by interfacial dislocations, c) Fast Fourier Transformation (FFT) of image (b), which shows the Burgers orientation relationship: $(0001)_\alpha$ // $(011)_\beta$ and $[2\text{-}1\text{-}10]_\alpha$ // $[1\text{-}11]_\beta$ between the precipitates and the matrix.

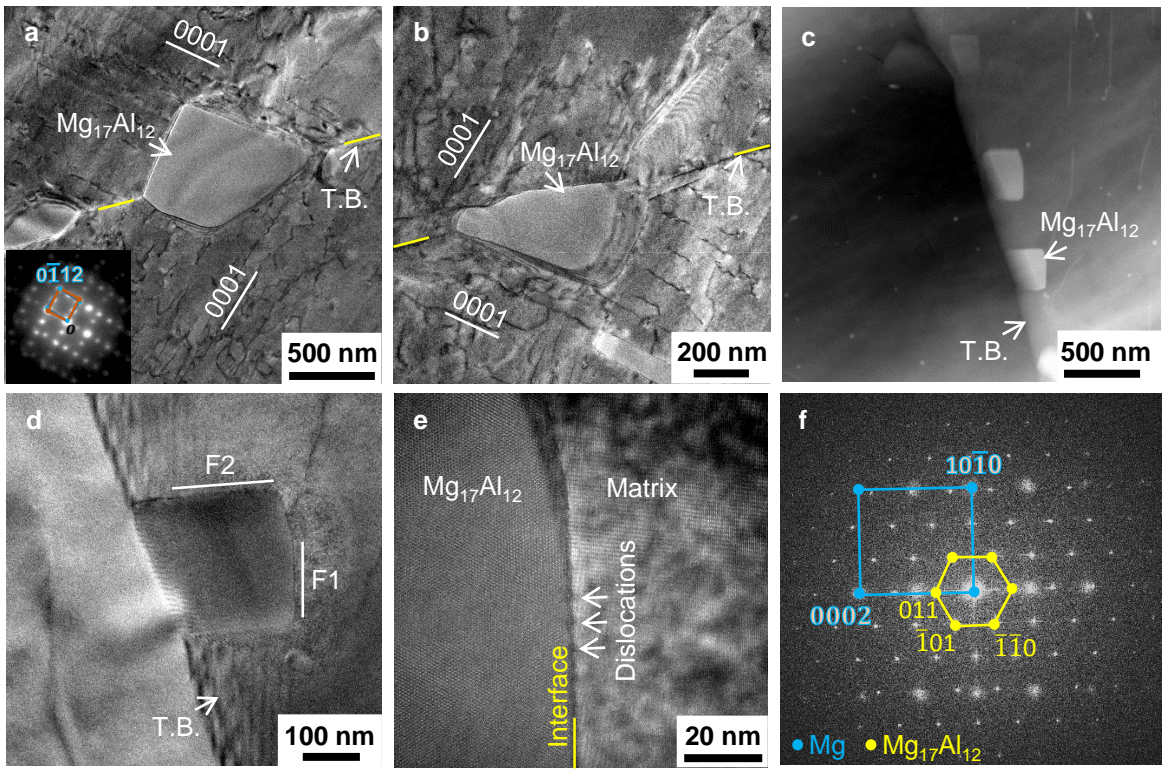

Figure 3. TEM images of β-Mg$_{17}$Al$_{12}$ precipitates along the twin boundary in Mg-Al-Zn alloy aged at 300°C for 2h. a) A precipitate across the twin boundary viewed along [2-1-10]$_α$ zone axis with twin boundary (T. B.) edge-on, and the inset are the diffraction pattern across the twin, which shows {0-112}$_α$ type twin, b) A precipitate nearly in one side of twin boundary viewed along [2-1-10]$_α$ zone axis, c) HAADF-STEM image of two precipitates with their interface edge-on along [-12-10]$_α$, d) enlarged view of one of the precipitate in figure (c), and the Facet 1 (F1) and Facet 2 (F2) are edge-on, c) High resolution image around F1, and the periodic strain contrasts on the interface are caused by interfacial dislocations, f) Fast Fourier Transformation (FFT) of image (e), which shows the Burgers orientation relationship.

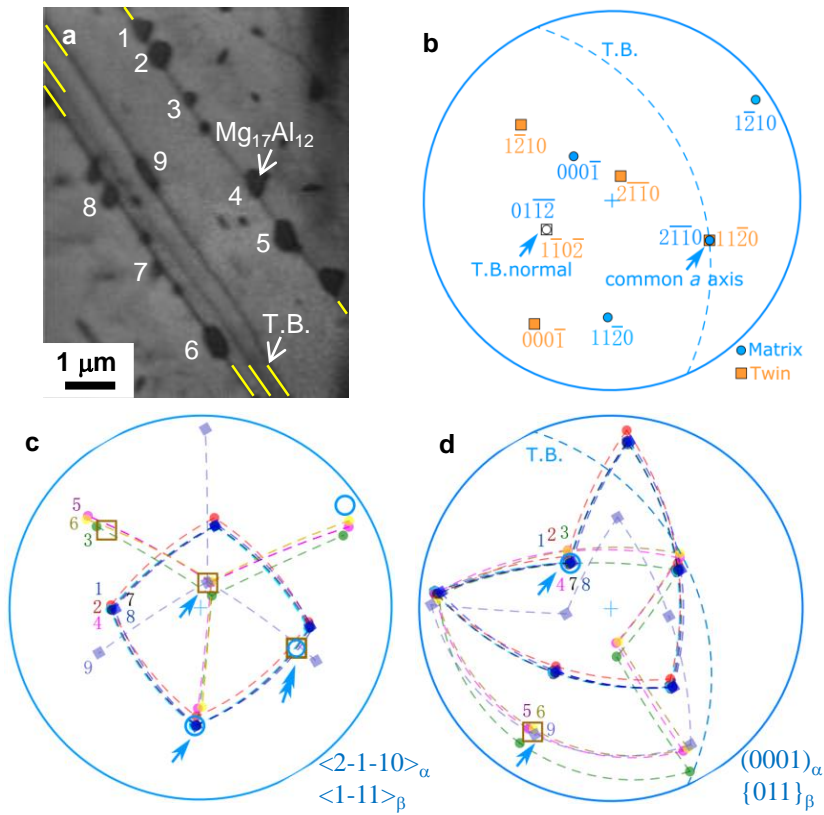

Figure 4. Statistic study of the OR between the matrix and the precipitates along the twin boundary by transmission EBSD (t-EBSD). a) Band contrast (BC) mapping, and the numbers in the figure indicates the precipitates for further analysis, b) pole figure for $\{0\text{-}112\}_\alpha$ type twin in figure (a), c) $<1\text{-}11>_\beta$ pole figure for the precipitates numbered in figure (a), and the $<2\text{-}1\text{-}10>_\alpha$ directions in matrix are also superimposed, d) $\{011\}_\beta$ pole figure of the precipitates numbered in Figure (a) and the $\{0001\}_\alpha$ planes are also superimposed. Blue circles indicate the indices in matrix while orange squares are for those in twin. For $<1\text{-}11>_\beta$ pole figure, there are four $<1\text{-}11>_\beta$ directions for one precipitate and they are connected with big circles for easier read, while in $\{011\}_\beta$ pole figure, there are six poles of $\{011\}_\beta$ planes for one precipitate, and they are also connected with big circles.

Table 1. Six variants of β-$Mg_{17}Al_{12}$ precipitates at Burgers orientation relationship (OR) in Mg matrix. The misorientation between different variants is shown in third column, and θ is the deviation angle from the Burgers OR with the other side of twin boundary $(0\text{-}112)_\alpha$.

| No. | OR | Mis. | θ |
|---|---|---|---|
| V1 | $(0001)_{Mg} // (011)_\beta$<br>$[2\text{-}1\text{-}10]_{Mg}//[1\text{-}11]_\beta$ | -- | 26.3° |
| V2 | $(0001)_{Mg} // (011)_\beta$<br>$[\text{-}1\text{-}120]_{Mg}//[1\text{-}11]_\beta$ | 60° | 8.2° |
| V3 | $(0001)_{Mg} // (011)_\beta$<br>$[\text{-}12\text{-}10]_{Mg}//[1\text{-}11]_\beta$ | 60° | 28.3° |
| V4 | $(0001)_{Mg} // (0\text{-}1\text{-}1)_\beta$<br>$[1\text{-}210]_{Mg}//[1\text{-}11]_\beta$ | 49.5° | 8.2° |
| V5 | $(0001)_{Mg} // (0\text{-}1\text{-}1)_\beta$<br>$[\text{-}2110]_{Mg}//[1\text{-}11]_\beta$ | 60° | 26.3° |
| V6 | $(0001)_{Mg} // (0\text{-}1\text{-}1)_\beta$<br>$[11\text{-}20]_{Mg}//[1\text{-}11]_\beta$ | 10.5° | 28.3° |

# Asymmetrical precipitation on {10-12} twin boundary in magnesium alloy

Xin-Fu Gu[1*], Man Wang[1], Zhang-Zhi Shi[1], Leng Chen[1], Ping Yang[1]

School of Materials Science and Engineering, University of Science and Technology Beijing, Beijing, 100083, China

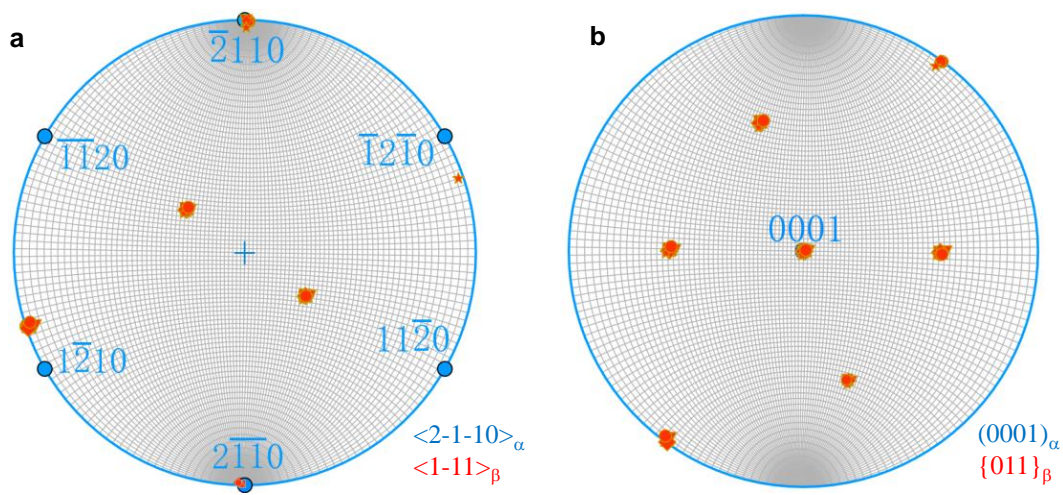

Figure S1. Statistic study of the OR between the matrix and the precipitates in matrix by transmission EBSD. a) <1-11>$_\beta$ pole figure superimposed with <2-1-10>$_\alpha$ pole figure of matrix, b) {011}$_\beta$ pole figure superimposed with {0001}$_\alpha$ pole figure of matrix.

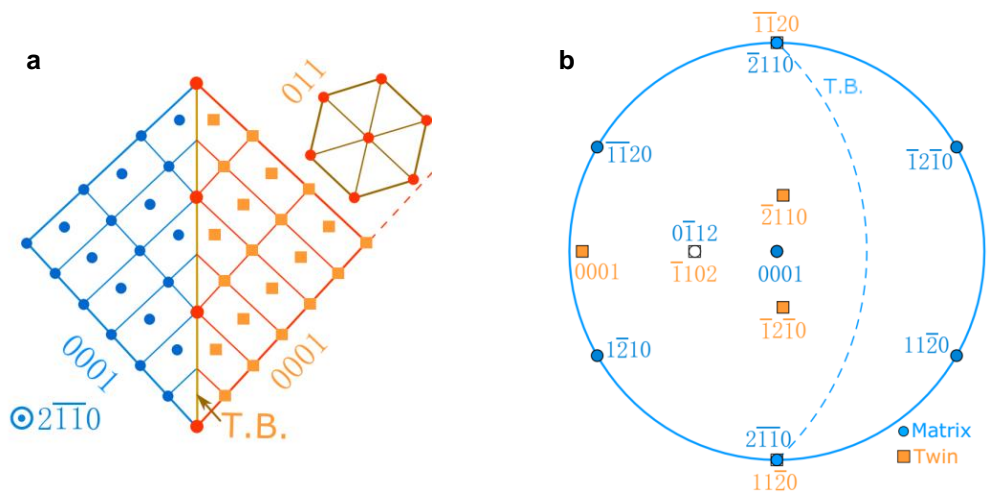

Figure S2. Twin relationship. a) atomic structure near the twin boundary viewed along [2-1-10]$_\alpha$, and the twin plane is (0-112) $_\alpha$, b) stereo-graphic projection of the twin relationship, where T.B. is short for twin boundary. (For interpretation of the references to color in this figure legend, the reader is referred to the web version of this article.)

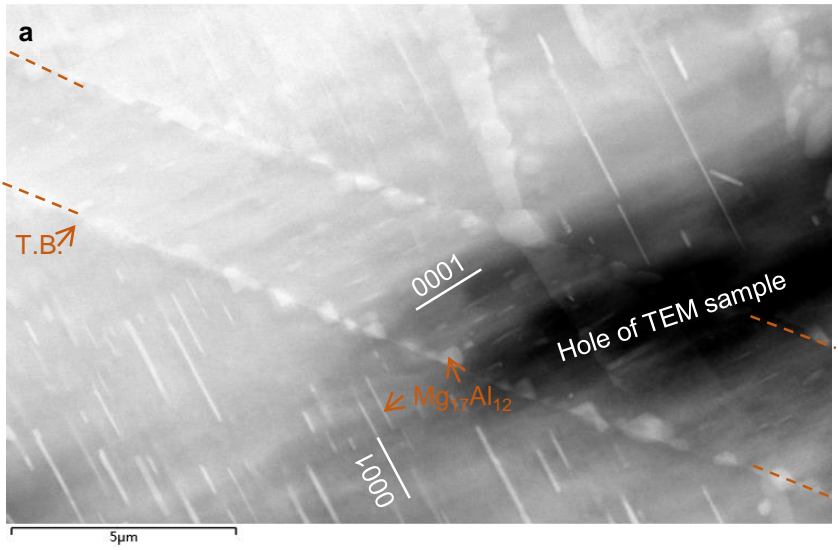
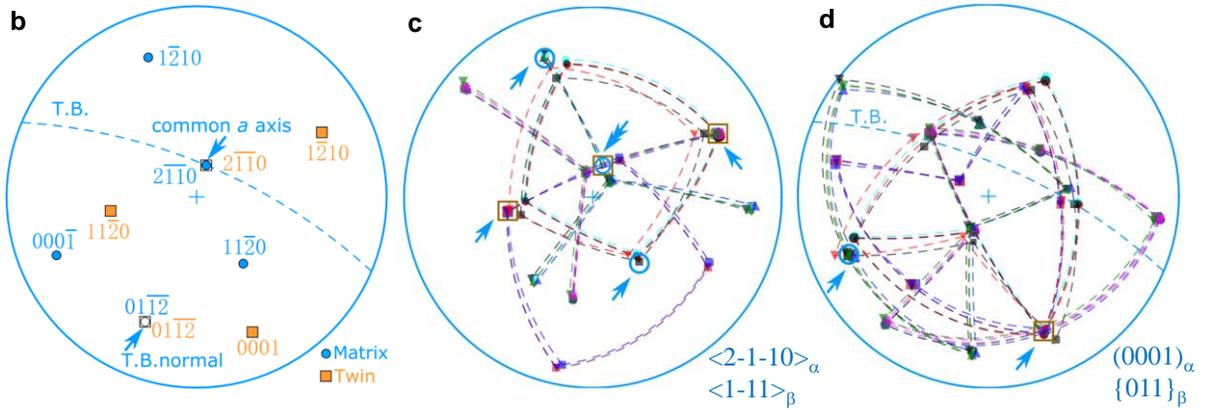

Figure S3. Statistic study of the OR between the matrix and the precipitates along the twin boundary by transmission EBSD. a) Microstructure imaging using forescatter diodes (FSD), and the precipitates along the twin boundary will be used for further analysis, b) pole figure for $\{10\text{-}12\}_\alpha$ type twin in figure (a), c) $<1\text{-}11>_\beta$ pole figure for the precipitates along twin boundary and the $<2\text{-}1\text{-}10>_\alpha$ directions in matrix are also superimposed, d) $\{011\}_\beta$ pole figure of the precipitates along twin boundary and the $\{0001\}_\alpha$ planes are also superimposed. Blue circles indicate the indices in matrix while orange squares are for those in twin.

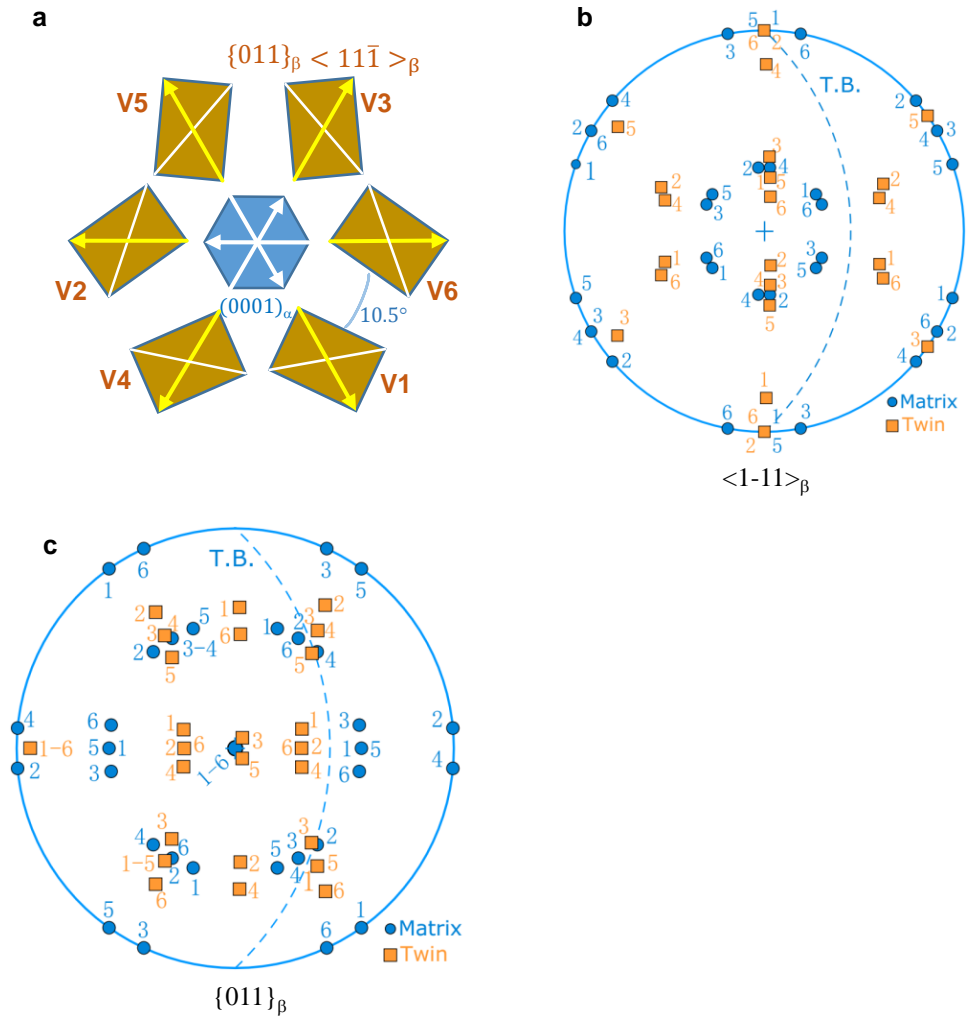

Figure S4. a) Six possible variants of exact Burgers orientation relationship between precipitates and Mg matrix. b) <1-11>$_\beta$ pole figure of the six variants of twin and matrix, respectively with different color, c) {011}$_\beta$ pole figure of the six variants of twin and matrix, respectively. The blue color is for the variants in matrix, while the orange color is for these in twin. The number in the pole figure indicates the variant number in Figure (a).